\documentclass[11pt]{article}
\newcommand{\be}{\begin{equation}}
\newcommand{\ee}{\end{equation}}
\newcommand{\bea}{\begin{eqnarray}}
\newcommand{\eea}{\end{eqnarray}}

\newcommand{\sn}{{\rm sn}}
\newcommand{\cn}{{\rm cn}}
\newcommand{\dn}{{\rm dn}}

\begin{document}
~\hfill{\footnotesize SUNYB/06-1, IOP-BBSR/06-2}
\vspace{.5in}
\begin{center}
{\LARGE {\bf Complex Periodic Potentials\\ 
with a Finite 
Number of Band Gaps}}
\end{center}
\vspace{.27in}
\begin{center}
{\Large{\bf  \mbox{Avinash Khare}
 }}\\
\noindent
{\large Institute of Physics, Sachivalaya Marg, Bhubaneswar 751005, India}
\end{center}
\begin{center}
{\Large{\bf  \mbox{Uday Sukhatme}
 }}\\
\noindent
{\large Department of Physics, State University of New York at Buffalo, Buffalo, NY 14260, U.S.A. }\\
\end{center}
\vspace{.7in}
{\bf {Abstract:}}  
We obtain several new results 
for the complex generalized associated 
Lam\'e potential 
$$ 
\hspace{-1.5in}V(x)=a(a+1)m~\sn^2(y,m)+b(b+1)m~{\sn^2 (y+K(m),m)}
$$
$$\hspace{0.5in}+f(f+1)m~{\sn^2 (y+K(m)+iK'(m),m)}+g(g+1)m~{\sn^2 (y+iK'(m),m)}~,
$$
where $y \equiv x-\frac{K(m)}{2}-\frac{iK'(m)}{2}$, $\sn(y,m)$ is a Jacobi elliptic function 
with modulus parameter $m$, and there are four real parameters $a,b,f,g$. 
First, we derive two new duality relations which, when coupled with a 
previously obtained duality relation, permit us to relate 
the band edge eigenstates of the 24 potentials obtained by permutations of the
four parameters $a,b,f,g$. Second, we pose and answer the question:
how many independent potentials are there with
a finite number ``$a$" of band gaps when $a,b,f,g$ are integers? 
For these potentials, we clarify the nature of 
the band edge eigenfunctions. We also obtain several analytic results when at least one of
the four parameters is a half-integer. As a by-product, we also
obtain new solutions of Heun's differential equation.

\newpage

\section{Introduction.}

In a recent paper \cite{ks7}, hereafter referred to as I, we discussed the
generalized associated Lam\'e (GAL) potential given by
\bea\label{0}
\hat{V}(x)&=&a(a+1)m~\sn^2(x,m)+b(b+1)m~{\sn^2 (x+K(m),m)}
 \nonumber \\
&&~~~+f(f+1)m~ {\sn^2 (x+K(m)+iK'(m),m)}+g(g+1)m~{\sn^2 (x+iK'(m),m)} \nonumber \\
&=& a(a+1)m~\sn^2(x,m)+b(b+1)m~\frac{\cn^2 (x,m)}{\dn^2 (x,m)} 
+f(f+1) \frac{\dn^2 (x,m)}{\cn^2(x,m)}+g(g+1)\frac{1}{\sn^2 (x,m)}~, \nonumber \\ 
\eea
which involves four real parameters $a,b,f,g$. Here, $\sn ~(x,m)$, 
$\cn ~(x,m)$, $\dn ~(x,m)$ are Jacobi elliptic 
functions with elliptic
modulus parameter $m$ $( 0\leq m \leq 1)$. They are doubly periodic
functions with periods $[4K(m), i2K'(m)]$, $[4K(m), 2K(m)+i2K'(m)]$,
$[2K(m), i4K'(m)]$ respectively \cite{abr}, where 
$K(m) \equiv \int_0^{\pi/2} d\theta [1-m\sin^2 \theta]^{-1/2}$ 
denotes the complete elliptic integral of the first kind, and
$K'(m)\equiv K(1-m)$. From now on, unless essential,  
we will not explicitly display the modulus parameter $m$ as an 
argument of Jacobi
elliptic functions. It may be noted here that the four terms in the 
GAL potential (\ref{0}) correspond 
to complex translations of the independent variable $x$ by
$0, K(m), K(m)+iK'(m), iK'(m)$.  Although the GAL potential is real, it
does have singularities on the real axis coming from the zeros of the Jacobi
elliptic functions $\sn ~(x)$ and $\cn ~(x)$ in the last two 
terms. 
One way to avoid these singularities is to make a 
complex change of variables $y=ix+\beta$, with $\beta$ being an arbitrary 
real, non-zero constant, chosen
so as to avoid the singularities arising from the zeros of Jacobi elliptic 
functions on
the real axis \cite{bed}. This procedure was used in I, and we studied many 
of the properties 
of the resulting PT-invariant complex periodic potential \cite{ks5}. 
However, consistent
with the practice in the mathematics literature, in this paper we use 
an alternative approach of avoiding singularities by simply 
translating the independent variable $x$ by an arbitrary non-zero amount 
in the complex plane. In fact, for simplicity, from
symmetry considerations, we take the translated variable to be
$x-\frac{K(m)}{2}-\frac{iK'(m)}{2}$. Note that the potential still is a
PT-invariant complex periodic potential. One very important consequence of
this choice is that the energy eigenvalues which we shall obtain here will 
be opposite in sign from the values obtained in I. This point
should be kept in mind while comparing the results from 
the two papers. Thus explicitly, we consider the potential 
\bea\label{1}
V(x)&=&\hat{V}(y)\nonumber \\
&=&a(a+1)m~\sn^2(y,m)+b(b+1)m~{\sn^2 (y+K(m),m)}
 \nonumber \\
&&+f(f+1)m~ {\sn^2 (y+K(m)+iK'(m),m)}+g(g+1)m~{\sn^2 (y+iK'(m),m)} \nonumber \\
&=& a(a+1)m~\sn^2(y,m)+b(b+1)m~\frac{\cn^2 (y,m)}{\dn^2 (y,m)} 
+f(f+1) \frac{\dn^2 (y,m)}{\cn^2(y,m)} 
+g(g+1)\frac{1}{\sn^2 (y,m)} \nonumber \\
&\equiv & [a,b,f,g]~, 
\eea 
where 
\be\label{1z}
y=x-\frac{K(m)}{2}-\frac{iK'(m)}{2}~.
\ee
It may be noted here that in I we used the notation
$[a(a+1),b(b+1),f(f+1),g(g+1)]$ to denote this potential.  
However, for consistency with the prevailing practice in the mathematics literature, 
in this paper we use the notation $[a,b,f,g]$. 
In this notation, ordinary Lam\'e potentials are denoted by
$[a,0,0,0]$, and associated Lam\'e (AL) potentials 
are denoted by
$[a,b,0,0]$. Note that the potential (\ref{1}) remains unchanged when any one
or more of the parameters $a,b,f,g$ change to $-a-1,-b-1,-f-1,-g-1$ 
respectively.

There is one important point which
permits us to construct many supersymmetric partner
potentials corresponding to a given potential \cite{ks6}. Although this point was 
previously made in I, it is worth restating here. 
Normally, in supersymmetric quantum mechanics \cite{cks}, 
given a potential $V_{-}(x)$, 
the ground state wave function $\psi_0(x)$ is used to construct the 
superpotential $W(x) = -\psi_0'(x)/\psi_0(x)$, 
which then yields the supersymmetric (SUSY) 
partner potential $V_{+}(x)=W^2+W'$. If one 
uses any excited state wave function 
$\psi(x)$ of $V_{-}(x)$ to construct a superpotential $W(x)$, then the original 
potential $V_{-}(x)$ is recovered correctly
(by construction), but 
the corresponding partner potential $V_{+}(x)$ turns
out to be singular on the real $x$-axis 
due to the zeros of the excited state 
wave function $\psi(x)$. However, for the complex potential (\ref{1}), 
the singularities are not
on the real axis \cite{rl}. Thus in this case one could also use 
any of the excited state wave functions to obtain the superpotential
$W(x)$ and hence discover several supersymmetric partner potentials 
with the same energy spectrum.

In I we showed that several GAL potentials with specific
integer values of $a,b,f,g$ have a finite
number of band gaps. 
Further, looking at the symmetry of
these potentials, we conjectured that all GAL 
potentials with integer values of $a,b,f,g$ also 
have a finite number of band gaps.
Some results of this type are available in the mathematics literature, and it is 
worthwhile to present a brief review of what is already
known about the GAL potential.

The GAL potential (\ref{1}), expressed in terms of Weierstrass
functions, was discussed in a brief note by Darboux in 1882 \cite{dar}, as well 
as in two subsequent articles in 1914 and 1915. He mentioned
that some results had already been presented by Hermite in 1872 in unpublished 
lectures at Ecole Polytechnique. In 1883, Sparre \cite{spa}
wrote two long papers on the GAL potential expressed in terms of Jacobi
elliptic functions. Unfortunately, the mathematics community was largely unaware
of these papers, and the GAL potential was ``rediscovered" by Treibich and
Verdier in 1990 \cite{tv}. 
In the current mathematics literature, the GAL potential, expressed in terms of
Weierstrass functions, is known as the Treibich-Verdier potential.
Several workers have shown that when $a,b,f,g$ are integers, these
potentials have a finite number of band gaps. In particular, the number of
band gaps $p$ is given by \cite{gw}
\be\label{3}
p={{1} \over {2}}~ {\rm max}\bigg [2~ {\rm max}~[a,b,f,g], ~1+N- \bigg \{1+(-1)^{N} \bigg \} 
\bigg \{{\rm min}~[a,b,f,g]
+ \frac{1}{2} \bigg \} \bigg ] ~,
\ee
where $N=a+b+f+g$.
It was also shown that finite band gap potentials are solutions of
higher order KdV equations \cite{nov}. In recent times, several people have also
discussed the connection between Heun's equation and the Treibich-Verdier
potential \cite{tak}. Finally, some authors have studied even more general 
potentials with a finite number of band gaps \cite{tre}. 

In I using supersymmetry we showed that the band edge eigenvalues of
the Lam\'e and the AL potentials $[2,0,0,0],[3,0,0,0],[a,(a-3),0,0]$ 
are the same as those of the GAL potentials 
$[1,1,1,0],[2,1,1,1],[a,a-1,1,0]$ respectively.
We had also conjectured in I that for integer $a,b$, 
Lam\'e  and AL potentials have the same band edge eigenvalues
as some GAL potentials - the explicit relationship being:
\bea\label{2.16}
&&[a,0,0,0] \equiv \bigg [\frac{a}{2},\frac{a}{2},\frac{a}{2},\frac{a-2}{2}\bigg ]~,~
~~a= {\rm even~ integer}~, \nonumber \\
&&[a,0,0,0] \equiv \bigg [\frac{a+1}{2},\frac{a-1}{2},\frac{a-1}{2},\frac{a-1}{2}\bigg ]~,~
~~a= {\rm odd~ integer}~, \nonumber \\
&&[a,b,0,0] \equiv \bigg [\frac{a+b}{2},\frac{a+b}{2},\frac{a-b}{2},\frac{a-b-2}{2}\bigg ]~,~
~~a+b= {\rm even~ integer}~, \nonumber \\
&&[a,b,0,0] \equiv
\bigg [\frac{a+b+1}{2},\frac{a+b-1}{2},\frac{a-b-1}{2},\frac{a-b-1}{2}\bigg ]~,~
~~a+b= {\rm odd ~integer}~.
\eea
Note that we are using the notation ``$\equiv$" to denote ``same band edge eigenvalues", 
but not identical potentials. It is interesting to note that recently, 
Takemura \cite{tak1} has verified the conjectures expressed in (\ref{2.16}). 
Further, more generally, he has proved that the GAL
potential $[a,b,f,g]$ for integer $a,b,f,g$ has the same band edge
eigenvalues as another GAL potential, with the
explicit relationship depending on whether $N \equiv a+b+f+g$ is an even or an odd
integer. If $N$ is an even integer, the relationship is
\be\label{2.17}
[a,b,f,g] \equiv 
\bigg [\frac{a+b+f-g}{2},\frac{a+b+g-f}{2},\frac{a+f+g-b}{2},
\frac{b+f+g-a}{2}\bigg ]~,
\ee
while if $N$ is an odd integer, then the relationship is
\be\label{2.18}
[a,b,f,g] \equiv 
\bigg [\frac{a+b+f+g+1}{2},\frac{a+b-f-g-1}{2},\frac{a+f-b-g-1}{2},
\frac{a+g-b-f-1}{2}\bigg ]~.
\ee
He has also shown that if $a,b,f,g$
are all half-integers and their sum is an even integer, then the
band edge energy eigenvalues of the potential $[a,b,f,g]$ 
are the same as two other GAL potentials
where also all four parameters are half-integers with their sum being
an even integer. In particular, the explicit relationships are
\bea\label{4.6}
&&\bigg [a=k+\frac{1}{2},b=l+\frac{1}{2},f=n+\frac{1}{2},g=p+\frac{1}{2}\bigg ] \nonumber \\
&&~~~\equiv \bigg [\frac{k+l+n+p+3}{2},\frac{k+l-n-p-1}{2},
\frac{k+n-l-p-1}{2},\frac{k+p-n-l-1}{2} \bigg ] \nonumber \\ 
&&~~~\equiv \bigg [\frac{k+l+n-p+1}{2},\frac{k+l+p-n+1}{2},\frac{k+n+p-l+1}{2},
\frac{l+n+p-k+1}{2} \bigg ]~.
\eea

The above mentioned results, when combined with 
our work in I, 
raise several questions. For example, 
while it is clear from the relation
(\ref{3}) that all the 24 potentials obtained by permutation of the four
parameters $a,b,f,g$ have the same number of band gaps, what is the precise
relationship between the band edge eigenstates of these 24 potentials? 
Second, a very interesting 
question is to ask how many independent
GAL potentials are there with say $a$ band gaps. Third, is it possible to
further generalize the results of Takemura \cite{tak1} ?
Besides, what is the nature of the
band edge eigenfunctions for a general GAL potential in case 
$a,b,f,g$ are all integers? 
Finally, in view of the connection between the GAL
potentials and Heun's equation \cite{spa,tak}, 
it is worth enquiring about the
implications of these results in the context of 
solutions to Heun's equation. 

In this paper, we address all the above-raised issues. Further, we also
consider GAL potentials in which at 
least one of the four parameters is a
half-integer while the other parameters are arbitrary numbers and show that 
both relations (\ref{2.17}) and (\ref{2.18}) are valid in all
these cases. As a consequence, we conjecture that the relations
(\ref{2.17}) and (\ref{2.18}) are in fact simultaneously valid even when 
all four parameters $a,b,f,g$ are integers with their band edge 
eigenvalues being the same as the quasi-exactly solvable (QES) 
eigenvalues of potentials where all
four parameters are half-integers (with their sum being an odd integer).
Further, knowing the QES eigenvalues of the GAL potentials $[a,1/2,f,g],
[a,b,1/2,g],[a,b,f,1/2]$ (where $a,b,f,g$ are arbitrary numbers), and
using the connection between the Schr\"odinger equation
for the GAL potentials and Heun's equation, we show that the 
corresponding 
eigenfunctions can be obtained 
by directly solving the algebraic form of Heun's equation.

The plan of this paper is as follows. In Sec. 2,  
we derive new duality relations and examine some
consequences. In particular, using these duality relations we
obtain the precise connection between the energy eigenvalues 
and eigenfunctions of the 24
potentials obtained by permuting the four parameters $a,b,f,g$. 
In Sec. 3, we discuss the GAL potential (\ref{1})  
in some detail when the four parameters $a,b,f,g$ are all integers.
First, 
we find various GAL potentials which are related to each other
in the sense that they 
have identical band edge energy eigenvalues. We also
find a large number of self-isospectral potentials as well as
self-dual but non-self-isospectral potentials. 
Using all these results, we obtain the number of independent
potentials with say $a$ band gaps. We also clarify the nature of
the band edge eigenfunctions for these potentials. 
In Sec. 4 we
discuss the GAL potential when either one or more of the four
parameters take half-integer values while the remaining parameters are
arbitrary. In this case, in general one expects to
obtain QES mid-band states, by which we mean any energy state lying inside 
an energy band and thus not a band edge. Quite remarkably,
by generalizing Takemura's results \cite{tak1}  we obtain GAL potentials
which
have the same band edge eigenvalues as the mid-band energy values of the 
potentials with one or more of the parameters being half-integers.
In Sec. 5 we discuss the GAL potential in case the parameters $a,b,f,g$ take
arbitrary values and obtain the corresponding GAL potentials
with the same band edge and mid-band energy values when $a+b+f+g$
[or any other combination obtained by replacing one or more of these 
parameters by $-a-1,-b-1,-f-1,-g-1$ respectively] is an even integer.
In Sec. 6 we discuss the implications of all these
results in the context of Heun's equation. In particular, we show that in
view of the connection between the different GAL potentials, 
given a periodic solution of Heun's equation, one immediately obtains four periodic and
three quasi-periodic solutions of 
the same equation. We also show 
that in many cases, knowing the QES energy values of a GAL potential, 
it is much easier to
solve the algebraic form of Heun's equation and obtain the corresponding
eigenfunctions of the GAL potential. 
Finally, in Sec. 7 we
summarize the results obtained in this paper and spell out some 
open problems.

\section{Duality Relations for GAL Potentials.}

In this paper our main focus is on the Schr\"odinger equation
\be\label{2.1}
-\frac{d^2}{dx^2}\psi(x)+\hat{V}(x)\psi(x)=E\psi(x)~,
\ee
where $\hat{V}(x)$ is the GAL potential given by eq. (\ref{0}), and we have 
chosen units with $\hbar=2m=1$. 
Displaying parameters more explicitly, eq. (\ref{2.1}) states that 
the potential $\hat{V}(a,b,f,g,m;x)$ has band edge eigenvalues $E(a,b,f,g,m)$ 
and eigenfunctions $\psi(a,b,f,g,m;x)$. Of course, this means that the translated potential 
$V(x) \equiv V(a,b,f,g,m;x)=\hat{V}(a,b,f,g,m;x-\frac{K(m)}{2}-\frac{iK'(m)}{2})$ given in eq. (\ref{1}) has the same eigenvalues $E(a,b,f,g,m)$ 
and translated eigenfunctions $\psi(a,b,f,g,m;x-\frac{K(m)}{2}-\frac{iK'(m)}{2})$. 

First we want to
show that the band edge eigenstates of the 24 potentials [obtained
via permutations of the 4 parameters, $a,b,f,g$ in $V(x)$] are all related, so that once the
band edge eigenstates of any one permutation are known, the complete band 
edge eigenstates of all 24
potentials are also known.
Actually, these relations are valid for both band edges as well as mid-band states. 

The first relation is simple - in view of the invariance of the
Schr\"odinger eq. (\ref{2.1}) under
the translation $x \rightarrow x+K(m)$ followed by the interchanges 
$a \leftrightarrow b$
and $f \leftrightarrow g$, one gets \cite{ks7}
\be\label{2.2} 
E(b,a,g,f,m) = E(a,b,f,g,m)~,~~\psi(b,a,g,f,m;x) ~\propto~
\psi(a,b,f,g,m;x+K(m))~.
\ee
Similarly, the translations $x \rightarrow x+iK'(m)$ and 
$x \rightarrow x+K(m)+iK'(m)$ followed by suitable interchanges of
parameters, yield 
\be\label{2.4}
E(g,f,b,a,m) = E(a,b,f,g,m)~,~~\psi(g,f,b,a,m; x) ~\propto~
\psi(a,b,f,g,m; x+iK'(m))~,
\ee
\be\label{2.3}
E(f,g,a,b,m) = E(a,b,f,g,m)~,~~\psi(f,g,a,b,m; x) ~\propto~
\psi(a,b,f,g,m; x+K(m)+iK'(m))~.
\ee

Thus, once we obtain the eigenvalues and eigenfunctions of a given
potential $[a,b,f,g]$, then
we immediately know the eigenvalues and eigenfunctions 
 of three other potentials:
$[b,a,g,f]$, $[g,f,b,a]$ and $[f,g,a,b]$. Note that relations (\ref{2.2}),
(\ref{2.4}), (\ref{2.3}) all involve the same modulus parameter $m$.

We now derive three remarkable duality relations, which connect the 
energy states of different GAL potentials, and involve changes in the 
modulus parameter from $m$ to $1-m$, $1/m$ and ${-m}/{(1-m)}$.

\noindent{\bf Duality Relation I:}
This was
already derived in I and is given by
\bea\label{2.5}
&&E(a,b,f,g,m)=[a(a+1)+b(b+1)+f(f+1)+g(g+1)]
-E(a,g,f,b,1-m)~, \nonumber \\
&&\psi(a,b,f,g,m; x) ~\propto~ \psi(a,g,f,b,1-m; ix+K'(m)+iK(m))~.
\eea

\noindent{\bf Duality Relation II:}
Using the formulas \cite{abr}
\be\label{2.6}
\sn(x,m)= \frac{1}{\sqrt{m}}\sn(\sqrt{m}x,\frac{1}{m})~~,~\cn(x,m)=\dn(\sqrt{m}x,\frac{1}{m})~,
~\dn(x,m)=\cn(\sqrt{m}x,\frac{1}{m})~,
\ee 
and redefining a new variable $z=\sqrt{m}x$, the Schr\"odinger eq. (\ref{2.1}) 
takes the form
\bea\label{2.7}
&&-\frac{d^2}{dz^2}\psi(z)+\bigg
[\frac{a(a+1)}{m}\sn^2(z,\frac{1}{m})+\frac{f(f+1)}{m}
\frac{\cn^2(z,\frac{1}{m})}{\dn^2(z,\frac{1}{m})}+b(b+1) 
\frac{\dn^2(z,\frac{1}{m})}{\cn^2(z,\frac{1}{m})} \nonumber \\
&&
~~~~~~~~~~~~~~~~~~~~~~~~~~~~~~+g(g+1) \frac{1}{\sn^2(z,\frac{1}{m})} \bigg ] \psi(z)
= \frac{E(m)}{m} \psi(z)~.
\eea
On comparing eqs. (\ref{2.1}) and (\ref{2.7}) we obtain the duality
relation
\be\label{2.8}
E(a,b,f,g,m)=m E (a,f,b,g,\frac{1}{m})~,~
\psi(a,b,f,g,m;x) \propto \psi (a,f,b,g,\frac{1}{m};\sqrt{m}x)~.
\ee

\noindent{\bf Duality Relation III:}
We again start from the Schr\"odinger equation (\ref{2.1}) and now 
use the formulas \cite{abr}
\bea\label{2.10}
&&\sn(x,m)= \frac{\sn[\sqrt{1-m}~x,\frac{-m}{1-m}]}{\sqrt{1-m}~\dn[\sqrt{1-m}~x,\frac{-m}{1-m}]}~,
\nonumber \\
&&\cn(x,m)= \frac{\cn[\sqrt{1-m}~x,\frac{-m}{1-m}]}{\dn[\sqrt{1-m}~x,\frac{-m}{1-m}]}~,
\nonumber \\
&&\dn(x,m)= \frac{1}{\dn[\sqrt{1-m}~x,\frac{-m}{1-m}]}~.
\eea
On defining a new variable $z=\sqrt{1-m}~x$, the Schr\"odinger eq. (\ref{2.1}) 
takes the form
\bea\label{2.11}
&&-\frac{d^2}{dz^2}\psi(z)+\bigg
[-\frac{m~b(b+1)}{1-m} \sn^2 (z,\frac{-m}{1-m})
-\frac{m~a(a+1)}{1-m} \frac{\cn^2 (z,\frac{-m}{1-m})}{\dn^2
(z,\frac{-m}{1-m})} 
+f(f+1) \frac{\dn^2 (z,\frac{-m}{1-m})}{\cn^2
(z,\frac{-m}{1-m})}  \nonumber \\ 
&& ~~~~~~~~~~~~~~~~~~~~~~~~~~~~~~~~~~~~~~~~~~~+g(g+1) \frac{1}{\sn^2
(z,\frac{-m}{1-m})}\bigg ] \psi(z)   \nonumber \\
&&= \frac{1}{1-m} E \psi(z)-\frac{m}{1-m} \bigg [a(a+1)+b(b+1)+f(f+1)+g(g+1) 
\bigg ] \psi (z). 
\eea
Comparing eqs. (\ref{2.1}) and (\ref{2.11}), one gets the duality
relation
\bea\label{2.12}
&&E(a,b,f,g,m)=(1-m) E (b,a,f,g,\frac{-m}{1-m})
+m \bigg [a(a+1)+b(b+1)+f(f+1)+g(g+1) \bigg ]~, \nonumber
\\
&&\psi(a,b,f,g,m;x) \propto \psi (b,a,f,g,\frac{-m}{1-m};\sqrt{1-m}~x)~.
\eea

Using the three duality relations [eqs. (\ref{2.5}),
(\ref{2.8}), (\ref{2.12})] along with the translation results [eqs. (\ref{2.2}),
(\ref{2.4}), (\ref{2.3})], it is easily shown
that once the eigenstates of a
given potential are known, we can
immediately obtain the energy eigenstates of all the 24 potentials
obtained by permuting the 4 parameters $a,b,f,g$. 
Hence, out of these 24 potentials, there is only one independent
potential and without loss of any generality, 
throughout this paper we only consider the potential
$[a,b,f,g]$ with $a \ge b \ge f \ge g $ (unless stated
otherwise). 

The duality relations are very powerful and have many interesting
consequences. For example, we find that
for arbitrary integer values of $a,f$, 
the potential $[a,0,f,0]$ has only a finite number of band gaps. 
This follows because, eq. (\ref{2.8}) gives
\be\label{2.9}
E(a,0,f,0,m)=m E(a,f,0,0,\frac{1}{m})~,
\ee
so that both the potentials $[a,f,0,0]$ and $[a,0,f,0]$ must have the same number of band edges.
Since one knows \cite{ks2} that the  AL potential $[a,f,0,0]$
has a finite number of band gaps, the same statement holds for the
potential $[a,0,f,0]$. 

\section{Independent GAL Potentials with $a$ Band Gaps [$a,b,f,g$ = integers].}
In this section, we want to answer the following interesting question: 
given a potential of the form $[a,b,f,g]$ with
$a \ge b \ge f \ge g \ge 0$,
how many independent GAL potentials are there with exactly 
$a$ band gaps?

To begin, for a given integer $a$, 
let us calculate the total number of 
possible GAL 
potentials with $a \ge b \ge f \ge g \ge 0$. The number of such
potentials is
\be\label{3.1x}
\sum_{b=0}^{a} \sum_{f=0}^{b} \sum_{g=0}^{f} 1= (a+1)(a+2)(a+3)/6~.
\ee

Now due to the Landen transformations \cite{abr,ks4},  a potential 
$[a,a,b,b]$ with $a \ne b$ is essentially the same as the 
potential $[a,b,0,0]$ and hence not really distinct. There 
are $a$ such potentials.
Similarly, the potential $[a,a,a,a]$ is
related to the potential $[a,0,0,0]$ and hence not distinct. This follows
from the relations \cite{abr}
\be\label{3.3x}
\sn^2 (x) = \frac{1-\cn(2x)}{1+\dn(2x)}~,~~m\sn^2 (x+iK'(m),m)
=\frac{1}{\sn^2(x,m)}~,
\ee
\be\label{3.4x}
\sn^2(x,m)+\sn^2(x+K(m),m)+\sn^2(x+iK'(m),m)+\sn^2(x+K(m)+iK'(m),m)
=4\sn^2(2x+iK'(m))~.
\ee
Thus the number of 
distinct GAL potentials of the form $[a,b,f,g]$ with
$a \ge b \ge f \ge g \ge 0$ is given by
\be\label{3.5x}
N_{dist} = a(a+1)(a+5)/6~.
\ee

In view of the formula (\ref{3}) for the number of band gaps, it is
obvious 
that these distinct potentials 
are of two types - there are those with exactly $a$
band gaps and those with more than $a$ band gaps. Using eqs. 
(\ref{2.17}) and (\ref{3}) 
it follows that in case $a+b+f+g$ is an even
integer and $a+g \ge  b+f$, then the potential has $a$
band gaps, while if $a+g < b+f$, then it has $(a+b+f-g)/2$ band gaps. 
Similarly, using eqs. (\ref{2.18}) and (\ref{3}) 
it follows that in case $a+b+f+g$ is an odd
integer and $a \ge b+f+g$, then the potential has $a$ 
band gaps, while if $a < b+g+f$, then 
it has $(a+b+f+g+1)/2$ band gaps. 

It is now straightforward to count the number of independent
GAL potentials
$[a,b,f,g]$ with $a$ band gaps and show that  
\bea\label{2.24}
&&N_{a}=\frac{1}{18}[a^3+9a^2+6a+2]~,~ ~ a=1 ~ (\rm mod ~3)~; \nonumber \\
&&N_{a}=\frac{1}{18}[a^3+9a^2+6a-2]~,~ ~a=2 ~ (\rm mod ~3)~; \nonumber \\
&&N_{a}=\frac{a}{18}[a^2+9a+6]~,~ ~ ~ a= 0 ~ (\rm mod ~3) ~.
\eea

Several comments are in order at this stage.

\begin{enumerate}

\item There is only 1 independent 
potential with one band gap but
there are 3 independent potentials with 2 band gaps and 7 independent
potentials with 3 band gaps. This implies \cite{nov} 
that while there is only
one independent KdV equation of 3rd order, there should be 3 such
equations of 5th order and 7 such equations of 7th order. It is worth 
pointing out that indeed there are 3 independent 
equations of 5th order called Sawada-Kotera and Kaup-Kupershmidt
equations. 
It may be interesting to explicitly obtain all
the 7 independent KdV equations of 7th order.

\item The number of independent
potentials of the form $[a,b,f,g]$ having more than $a$ band gaps
is obtained by subtracting eq. (\ref{2.24})) from eq. (\ref{3.5x}):
\bea\label{2.25}
&&N_{>a}=\frac{1}{18} (a+2)(2a^2+5a-1),~ ~ a=1 ~ (\rm mod~3)~; \nonumber \\
&&N_{>a}= \frac{1}{18} (a+1)(2a^2+7a+2),~ ~a=2 ~ (\rm mod ~3)~; \nonumber \\
&&N_{>a} = \frac{1}{18} a(a+3)(2a+3),~ ~a= 0 ~ (\rm mod ~3)~.
\eea
For example, for $a=2$,  there are 4 potentials of the form $[a,b,f,g]$ with 
more than 2 band gaps:
\be\label{2.25a}
N_{>2} = [2,2,2,0],~[2,2,2,1],~[2,2,1,0],~[2,1,1,1]~.
\ee
For $a=3$,
there are 9 potentials with more than 3 band gaps:
\be\label{2.25b}
N_{>3} = [3,3,3,0],~[3,3,3,1],~[3,3,3,2],~[3,3,2,0],
~[3,3,2,1],~[3,3,1,0],~[3,2,1,1],
[3,2,2,0],[3,2,2,2]~.
\ee
\item If $a+b+f+g$ is an even
integer and further if $a+g=b+f,$ then both sides of eq. (\ref{2.17})
are identical and the corresponding potential is
self-dual. Similarly, if $a+b+f+g$ is an odd integer and 
$a=b+f+g+1$, then both sides of eq. (\ref{2.18}) are identical, and 
one has a self-dual
potential. As an illustration, the AL potential $[a,a-1,0,0]$ is 
self-dual, a fact we already knew from ref. \cite{ks1}. But we now get a 
large number of additional self-dual GAL potentials, like  
$[2,1,1,0]$, $[4,2,2,0]$, and $[4,2,1,0]$, for example.
 
\item A related question is, out of the above self-dual potentials how
many are also self-isospectral \cite{ks1,ds}?
It is easy to see that a choice of eigenfunction yielding 
self-isospectral potentials is
\be\label{2.19}
\psi = \frac{\dn^{a}(x,m)\sn^{b}(x,m)}{\cn^{b}(x,m)}~,
\ee
and that the corresponding self-isospectral potential has the form
\be\label{2.20}
[a,a-1,b,b-1]~,~~ b > 0~;~~  [a,a-1,0,0]~,~~ b=0~.
\ee
The fact that the AL potentials of the form
$[a,a-1,0,0]$ are
self-isospectral was established many years ago \cite{ks1}. 
However, what is new
is the realization that the  potentials $[a,a-1,b,b-1]$
are also self-isospectral for arbitrary integer values of $a,b$. 
Some examples are
$[1,0,0,0]$, $[2,1,0,0]$, $[3,2,2,0]$, $[3,2,1,1]$, and $[3,2,0,0]$.   
Thus out of all the self-dual potentials, those 
which are of form (\ref{2.20}) are also self-isospectral,
while the rest are self-dual but not self-isospectral. 
The number of self-dual potentials $N_{sd}$ which are 
not self-isospectral is
\bea\label{2.20A}
&&N_{sd} = \frac{1}{3}(a-1)^2~,~~a=1,4,7,... ~;\nonumber \\
&&N_{sd} = \frac{1}{3}a(a-2)~,~~a=2,3,5,6,... ~,
\eea
whereas 
the number of self-isospectral potentials is 
$N_{si}=a$.
As an illustration, the self-isospectral potentials with two band
gaps are $[2,1,0,0]$ and $[2,1,1,0]$ while there are no 
self-dual, non-self-isospectral potentials with two band gaps.
On the other hand, the self-isospectral potentials with three band gaps
are $[3,2,2,1]$, $[3,2,1,0]$, and $[3,2,0,0]$, while the only self-dual,
non-self-isospectral potential with three band gaps is $[3,1,1,0]$.

\item Clearly all potentials with $a+g > b+f$ or $a > b+f+g+1$ depending
on if $a+b+f+g$ is an odd or an even integer, have 
partner potentials as given by eqs. (\ref{2.17}) or (\ref{2.18})
respectively. Now out of these, some are SUSY partner potentials
while the rest are merely partner potentials. So let us count both types
of potentials. Now if two GAL potentials are SUSY partners, then one of
their 
eigenfunctions must be related to each other by $\psi_{II}=\psi_{I}^{-1}$.
Further, these two eigenfunctions must have the form
\be\label{2.20a}
\dn^{\alpha}(x)\cn^{\beta}(x)\sn^{\gamma}(x)~.
\ee
For such eigenfunctions, we know from I that $b=-\alpha,f=-\beta,g=-\gamma$ and
$a+b+f+g=0$. From here it is easy to show that the potential $[a,b,f,g]$
with $a$ band gaps has a SUSY GAL partner provided either $a+g = b+f+2$ or
$a = b+f+g+3$ depending on if $a+b+f+g$ is an even or an odd integer
respectively. Hence, all
potentials of the form $[a,b,f,g]$ with $a$ band gaps and satisfying 
$a+g > b+f+2$ or $a > b+g+f+3$ have merely partner  
potentials of the form
(\ref{2.17}) or (\ref{2.18}) respectively. It is worth emphasizing that
while these partner potentials have the
same band edge eigenvalues, none of
them are SUSY partner potentials. 
For example, $[2,0,0,0]$ and $[1,1,1,0]$ are SUSY partner GAL potentials 
with two band gaps. Similarly $[3,1,0,0]$ and $[2,2,1,0]$ are SUSY partner
GAL potentials with three band gaps.
On the other hand, $[4,0,0,0]$  and $[2,2,2,1]$ are merely partner potentials
with four band gaps. 

\item One can count the number of potentials ($N_{su}$) of the form
$[a,b,f,g]$ with $a$ band gaps 
having another GAL potential as its SUSY partner and it is
easy to show that
\bea\label{2.22}
&&N_{su}=\frac{1}{3}[a^2-1]~,~~a \ne 0 ~(\rm mod ~3)~; \nonumber \\
&&N_{su} =\frac{1}{3} a^2~,~~a= 0 ~(\rm mod~ 3) ~.
\eea

Finally, it is not difficult to show that the number of potentials
of the form $[a,b,f,g]$ with $a$ band gaps and having 
merely  a (non-SUSY)
partner potential of the form as given by eqs. (\ref{2.17}) and 
(\ref{2.18}) respectively is given by
\bea\label{2.23}
&&N_{nsu}=\frac{1}{18}(a-1)(a^2-2a-2)~,~ ~ a=1 ~ (\rm mod ~3)~; \nonumber \\
&&N_{nsu}=\frac{1}{18}(a-2)(a^2-a-2)~, ~ ~ a=2 ~ (\rm mod ~3)~; \nonumber \\
&&N_{nsu}= \frac{1}{18} a^2(a-3)~,~ ~ a= 0 ~ (\rm mod ~3) ~. 
\eea

\item On adding the number of self-dual (but non-self-isospectral), 
self-isospectral, SUSY partner
and (non-SUSY) merely partner potentials as given by eqs. (\ref{2.20}) to
(\ref{2.23}) respectively, as expected, 
we find that the number of independent
potentials with $a$ band gaps is as given by eq. (\ref{2.24}).

\item One obvious interesting question is whether there are non-GAL potentials
with a finite number of band gaps. The answer to this
question is yes. In particular, since the general form of the
eigenfunction for any GAL potential is of the form \cite{ks7}
\be\label{1x}
\psi_{GAL}(x) = \dn^{-b}(x)\cn^{-f}(x)\sn^{-g}(x) \sum_{k=0}^{N} A_k
\sn^{2k}(x)~,
\ee
it follows that non-GAL potentials of the form
\be\label{1y}
V_{+}(x)=V_{GAL} (x) -2\frac{d^2}{dx^2} \ln \psi_{GAL} (x)~,
\ee
are also finite gap potentials. In this context it is worth mentioning
that some people \cite{tre} have recently 
obtained potentials
with a finite number of band gaps 
which are more general than the GAL potentials. It is not clear 
if those potentials and the potentials (\ref{1y})
have any overlap.

\end{enumerate}

\subsection{The Nature of Band Edge Eigenstates}

In I we showed that if $a+b+f+g=2n$, then $n+1$ QES states can be
obtained for GAL potentials (\ref{1}) and they are of the form
given in eq. (\ref{1x}).
On using this key result as well as the fact that the GAL potential
$[a,b,f,g]$ remains unchanged when any one (or more) of the four
parameters $a,b,f,g$ is changed to 
$-a-1,-b-1,-f-1,-g-1$ respectively, it is easy to specify the nature of
the band edge eigenfunctions for any potential $[a,b,f,g]$ with
$a$ band gaps. The nature as well as the number of eigenstates crucially
depend on whether $a+b+f+g$ is an even or an odd integer - so we 
will discuss these situations separately.

\noindent{\bf $a+b+f+g$ = even integer:}

\noindent In this case it is easy to show that one has
$(a+b+f+g+2)/2$ eigenstates of the form
\be\label{22.1}
\sn^{-g}(x)\cn^{-f}(x)\dn^{-b}(x) F_{(a+b+f+g)/2} [\sn^2(x)]~, 
\ee
$(a+b-f-g)/2$ eigenstates of the form
\be\label{22.2}
\sn^{g+1}(x)\cn^{f+1}(x)\dn^{-b}(x) F_{(a+b-f-g-2)/2} [\sn^2(x)]~, 
\ee
$(a+f-b-g)/2$ eigenstates of the form
\be\label{22.3}
\sn^{g+1}(x)\cn^{-f}(x)\dn^{b+1}(x) F_{(a+f-b-g-2)/2} [\sn^2(x)]~, 
\ee
 and $(a+g-b-f)/2$ eigenstates of the form
\be\label{22.4}
\sn^{-g}(x)\cn^{f+1}(x)\dn^{b+1}(x) F_{(a+g-b-f-2)/2} [\sn^2(x)]~. 
\ee
If instead $b+g > a+f$, then one has $(b+f-a-g)/2$ eigenstates of the
form
\be\label{22.4a}
\sn^{g+1}(x)\cn^{-f}(x)\dn^{-b}(x) F_{(b+f-a-g-2)/2} [\sn^2(x)]~. 
\ee

\noindent{\bf $a+b+f+g$ = odd integer:}

\noindent In this case it is easy to show that one has
$(a+b+f-g+1)/2$ eigenstates of the form
\be\label{22.5}
\sn^{g+1}(x)\cn^{-f}(x)\dn^{-b}(x) F_{(a+b+f-g-1)/2} [\sn^2(x)]~, 
\ee
$(a+b+g-f+1)/2$ eigenstates of the form
\be\label{22.6}
\sn^{-g}(x)\cn^{f+1}(x)\dn^{-b}(x) F_{(a+b+g-f-1)/2} [\sn^2(x)]~, 
\ee
$(a+f+g-b+1)/2$ eigenstates of the form
\be\label{22.7}
\sn^{-g}(x)\cn^{-f}(x)\dn^{b+1}(x) F_{(a+f+g-b-1)/2} [\sn^2(x)]~, 
\ee
$(a-b-f-g-1)/2$ eigenstates of the form
\be\label{22.8}
\sn^{g+1}(x)\cn^{f+1}(x)\dn^{b+1}(x) F_{(a-b-f-g-3)/2} [\sn^2(x)]~, 
\ee
If instead $b+f+g>a-1$ then one has $(b+f+g-a+1)/2$ 
eigenstates of the form
\be\label{22.8a}
\sn^{-g}(x)\cn^{-f}(x)\dn^{-b}(x) F_{(b+f+g-a-1)/2} [\sn^2(x)]~, 
\ee
Here $F_{n}[\sn^2(x)]$ denotes a polynomial of order $n$ in $\sn^2(x)$. 
It is worth pointing out that not only 
the QES band edge eigenvalues are identical for the two partner potentials 
 as given either by eq. (\ref{2.17}) 
or by (\ref{2.18}), even
the nature of the band edge eigenfunctions in the two cases is also similar. 
For example, the potential
$[(a+b+f-g)/2,(a+b+g-f)/2,(a+f+g-b)/2,(b+f+g-a)/2]$
as given by eq. (\ref{2.17}) has the same band edge eigenvalues
and further the corresponding
eigenfunction is simply obtained 
from eqs. (\ref{22.1}) to (\ref{22.4a}) 
by replacing $a,b,f,g$ with
$(a+b+f-g)/2,(a+b+g-f)/2,(a+f+g-b)/2,(b+f+g-a)/2$ respectively.
Exactly the same is also true about the equivalent potential
given by eq. (\ref{2.18}) in case $a+b+f+g$ is an odd integer and
the corresponding eigenfunctions are exactly as given by eqs. (\ref{22.5})
to (\ref{22.8a}) but with the replacement of $a,b,f,g$ by
$(a+b+f+g+1)/2,(a+b-f-g-1)/2,(a+f-b-g-1)/2,(a+g-b-f-1)/2$ respectively. 
It follows from here that irrespective of whether $a+b+f+g$ is an odd or
an even integer, there are precisely $a$ bound bands, same $a$
number of band gaps and $2a+1$ number of band edges all of which are
analytically known in principle, beyond which there is a continuum band
extending up to $E=\infty$. Further, irrespective of whether $a+b+f+g$ is an
odd or an even integer, if 
$a+b$ is an even (odd) integer, then there are
$a+b+1$ band edges of period $2K (4K)$ and $a-b$ band edges of period 
$4K (2K)$. 

Thus in general the band structure of the GAL potentials is
unusual in that if $a+b$ is an even (odd) integer, then $b$ band gaps 
of period $4K (2K)$ must be of zero width, i.e. there must be $b$
doubly degenerate states of period $4K (2K)$. 
Unfortunately, till today
we do not know either the eigenvalue or the nature of the eigenfunction
of even one of these doubly degenerate states. 
One exception is the case of pure Lam\'e 
(and their GAL partners) potentials, i.e. when $b=f=g=0$, as in that case
depending on if $a$ is even or odd integer, one has $a+1$ band edges of
period $2K (4K)$ and $a$ band edges of period $4K (2K)$ and the band
structure is normal one, with no doubly degenerate states.

As an illustration, consider the GAL potentials with two band gaps. 
As seen above, 
there are 3 distinct potentials with 2 band gaps out of which we have
already discussed the band structure of the two potentials 
$[2,0,0,0]$ and $[2,1,0,0]$ \cite{ks2,ks1}. 
Thus it would be interesting to know the
band edges and the band structure of the remaining 
potential with two band gaps, i.e. $[2,1,1,0]$. Since
$a+b=3$, it follows from the above discussion that in this case 
there must be 4 band edges of period $4K$ and 1 band edge of period
$2K$. Using Table 4 of I it is easily seen that the eigenstate with
period $2K$ is given by
\be\label{2.27}
\psi=\dn^2(x)\sn(x)\cn^{-1}(x)~,~~E =9m~,
\ee
while out of the 4 band edges of period $4K$, one eigenstate has the form
\be\label{2.28}
\psi=\dn^{-1}(x)\sn(x)\cn^{2}(x)~,~~E =9~.
\ee
The three other eigenstates of period $4K$ have the form
\be\label{2.29}
\psi=\dn^{-1}(x)\cn^{-1}(x)[A+B\sn^2(x)+D\sn^4(x)]~,
\ee
and the corresponding three eigenvalues satisfy the cubic equation
\be\label{2.30}
r^3+8(1+m)r^2+80mr+64m(1+m)=0~,~~E=-r+1+m~.
\ee
Further, it is clear that there must be one doubly degenerate state of
period $2K$ whose eigenvalue and eigenfunctions are not known
analytically.

\section{\bf GAL Potentials [with at least one parameter $a,b,f,g$ = half-integer].}
So far, we have discussed GAL potentials when all the four parameters $a,b,f,g$
take integer values. We have seen that these are problems with a finite
number of band gaps. We now consider the case when at least one of
the parameters $a,b,f,g$ is a half-integer. 
In general, all such problems have an infinite number 
of band gaps and one has only a few QES states.

\subsection{\bf $a$ = half-integer:}

As mentioned in the introduction, we  
conjecture that the relations
(\ref{2.17}) and (\ref{2.18}) are simultaneously valid when at least one
of the four parameters $a,b,f,g$ is a half-integer. 
In particular, in case $a=k+\frac{1}{2}$ and $b,f,g$ are arbitrary
numbers, we assert that $k+1$ QES energy values are identical for three
GAL potentials, that is
\bea\label{3.1}
&&[a=k+\frac{1}{2},b,f,g] \nonumber \\
&&\equiv \bigg[\frac{2(k+b+f+g)+3}{4},\frac{2(k+b-f-g)-1}{4},
\frac{2(k+f-b-g)-1}{4},\frac{2(k+g-b-f)-1}{4}\bigg] \nonumber \\ 
&& \equiv \bigg[\frac{2(k+b+f-g)+1}{4},
\frac{2(k+b+g-f)+1}{4},\frac{2(k+f+g-b)+1}{4},
\frac{2(b+f+g-k)-1}{4}\bigg]~.
\eea
This relation needs some clarification. 
What is being conjectured here is
that there are $k+1$ QES mid-band energy values of the potential
with $a=k+1/2$  
and $b,f,g$ being arbitrary numbers, which are the same as the band edge energy 
eigenvalues of the two other potentials given in (\ref{3.1}).
For these two potentials,
we can always obtain
$k+1$ QES band edges, since for both
of them the sum of the four parameters characterizing the potentials is
$2k$ \cite{ks7}. 

We have explicitly verified our  
conjecture in the following cases: (i) Lam\'e potentials with
$a=1/2,3/2$ (and $b=f=g=0$); (ii) AL potentials
\cite{ks2} with $a=1/2,3/2$, $b=1,2,3$ 
(and $f=g=0$); (iii) GAL potentials with $a=1/2,3/2$, and 
either $b$ or $f$ or $g$ is arbitrary while the remaining two parameters
take any integer values.
For example, we know that the mid-band state
of the $a=1/2$ Lam\'e potential is at $(1+m)/4$. Using eq.
(\ref{3.1}), we 
predict that both the potentials
$[\frac{3}{4},-\frac{1}{4},-\frac{1}{4},-\frac{1}{4}]$ and  
$[\frac{1}{4},\frac{1}{4},\frac{1}{4},-\frac{1}{4}]$ 
must have a QES band edge
eigenvalue at $E=(1+m)/4$. Using Table 4 of I it is easily checked
that this is indeed so, with the corresponding band edge eigenfunctions
being $\psi = [\dn(x)\cn(x)\sn(x)]^{1/4}$ and $\psi =\dn^{-1/4}(x)
\cn^{-1/4}(x)\sn^{3/4}(x)$ respectively. We have similarly checked the 
equivalence in all the above-mentioned cases.
For example, using the results \cite{ks2}
for the AL potential $[3/2,1,0,0]$, we find (and verify using Table 4 of
I) that the GAL potentials
$[7/4,3/4,-1/4,-1/4]$ and $[5/4,5/4,1/4,-1/4]$ have QES band edges at
$E=(13+5m)/4 \pm \sqrt{9-9m+m^2}$. The corresponding band edge
eigenfunctions are also easily written down. Similarly, using the results
in I for the mid-band states of the GAL potentials we verify that  
the 
potentials $[t/2+1,t/2-1,(1-t)/2,-(1+t)/2]$ and
$[(1+t)/2,-(1-t)/2,1-t/2,t/2]$ indeed have a QES band edge eigenvalue
$E=t^2+9m/4$, for any non-integer $t$. 

Thus one is fairly confident that the remarkable relationship 
(\ref{3.1}) is indeed
valid. We now turn around and use (\ref{3.1}) to predict new results for the
mid-band states of GAL potentials with half-integer values of $a$ in case
$b,f,g$ take arbitrary values.
As an illustration, we predict that the potential $[1/2,b,f,g],$ where
$b,f,g$ are arbitrary numbers 
must have a QES state at energy 
\be\label{3.4a}
E=(b+1/2)^2 +(f+1/2)^2 m~, 
\ee
while the potential $[3/2,b,f,g]$ with $b,f,g$ being arbitrary numbers
must have two QES states with energies
\bea\label{3.4}
&&E=[(b+1/2)^2+1]+[1+(f+1/2)^2]m \nonumber \\
&&\pm 2 \sqrt{[(b+1/2)+(f+1/2)m]^2-(b+f+g+3/2)(b+f-g+1/2)m}~.
\eea

Perhaps some clarification is required as to when the QES
state is a band edge and when it is a mid-band state. We believe that
only those states correspond to band edges for which $a+b+f+g$ [or any
other combination obtained by changing one or more of these parameters
to $-a-1,-b-1,-f-1,-g-1$ respectively] is equal to an even integer (including
zero). All other QES states should correspond to mid-band states.  
We thus believe that the QES energy values as given above for the 
potentials $[1/2,b,f,g]$ and $[3/2,b,f,g]$ with arbitrary $b,f,g$
are in most cases the energies for the mid-band states of these 
potentials. In Sec. 6 we shall obtain the QES eigenstates corresponding
to some of these QES eigenvalues by using the connection of the GAL
potential problem and Heun's equation.

We can also predict the QES energy values when 
$a=5/2$ and $b,f,g$ are arbitrary numbers, by computing the 
band edges of either of the two potentials given 
in (\ref{3.1}). In this way, we predict that the
mid-band energy values for the three QES states of the GAL potential 
$[5/2,b,f,g]$ are solutions of the cubic equation
\bea\label{3.5}
&&r^3+2[1+6b+(1+6f)m]r^2 +4[2(4b^2-1)+2(4f^2-1)m^2+(4b^2+4f^2-4g^2
+24bf \nonumber \\
&&+8b+8f-2g+3)m]r+8m[2b+1+(2f+1)m][4(b+f)^2-(2g+1)^2]=0~,
\eea
where $r=-E+(3/2-b)^2+(3/2-f)^2 m$. 
For the special cases (i) $b=f=g
=0$ as well as (ii) $b=1, f=g=0,$ it is easily checked that 
eq. (\ref{3.5}) agrees with well known results \cite{ks2}, 
thereby providing a powerful check on our calculations.
Generalization to higher half-integer values is straightforward (in
principle)  and it is easy to see that energy values for 
$a+1/2$ QES mid-band states can be predicted (at least in principle)
when $b,f,g$ are arbitrary numbers and $a$ is a half-integer.

\subsection{\bf $a,b$ = half-integers:}

Let us now discuss the case when both $a$ and $b$ are half-integers while $f$
and $g$ are arbitrary numbers using our conjecture that 
eqs. (\ref{2.17}) and (\ref{2.18}) are both  
simultaneously valid. In particular,   
for $a=k+\frac{1}{2},~b=l+\frac{1}{2}$ while $f,g$ are any numbers, 
we assert that the three potentials
\bea\label{4.1}
&&[a=k+\frac{1}{2},b=l+\frac{1}{2},f,g] \nonumber \\
&&\equiv \bigg[\frac{k+l+f+g+2}{2},\frac{k+l-f-g}{2},
\frac{k+f-l-g-1}{2},\frac{k+g-l-f-1}{2}\bigg] \nonumber \\ 
&& \equiv \bigg[\frac{k+l+f-g+1}{2},\frac{k+l+g-f+1}{2},\frac{k+f+g-l}{2},
\frac{l+f+g-k}{2}\bigg]~,
\eea
have the same $k+l+2$ QES energy values. 
This means 
that the QES  mid-band energy values of the potential with
$a=k+1/2,b=l+1/2$ are the 
same as the band edge eigenvalues of the two other potentials in (\ref{4.1}). 
For these two potentials,
we can obtain
$k+l+2=a+b+1$ QES states since, for both
of them, the sum of the four numbers characterizing the potentials is
either $2k$ or $2l$. This is possible because the GAL
potential $[a,b,f,g]$ remains unchanged when any one (or more) 
of the four parameters $a,b,f,g$ changes to $[-a-1,-b-1,-f-1,-g-1]$
respectively. 

As an illustration, consider the potential $[3/2,1/2,f,g]$ with $f,g$
being arbitrary numbers. It is then easily shown that this potential
must have 3 QES mid-band states at
\be\label{4.3a}
E_1 = 4+(g+1/2)^2 m~, 
\ee
\be\label{4.3b}
E_{2,3}=2+[(f+1/2)^2 +1]m 
\pm \sqrt{[2-(2f+1)m]^2 +[(2g+1)^2 -(2f-1)^2 m]}~.
\ee
Note that out of the $k+l+2$ QES states, the energies 
for the $k+1$ states can also be obtained by considering
the previous case of $a=k+1/2$ and $b,f,g$ arbitrary and putting 
$b=l+1/2$ at the end of the calculation. As an illustration, 
consider the case $[3/2,b,f,g]$ where $b$
is any arbitrary number.  As shown in the last section, there are
two QES energies given by eq. (\ref{3.4}).
On putting $b=1/2$ in eq. (\ref{3.4}) 
we find that the two QES energy values of
$[3/2,1/2,f,g]$ are precisely as given by eq. (\ref{4.3b}).
For the special case $f=g=0$, these eigenvalues 
agree with well-known results for the AL potential \cite{ks2}.
Similarly, consider the case $[5/2,b,0,0]$.
As shown in the previous subsection, 
there are three QES energy values as given by eq. (\ref{3.5}). 
On putting $b=1/2$, it is easily seen
that the three eigenvalues are $E=(1+9m/4),(1+25m/4),(9+m/4)$, 
in agreement with the eigenvalues obtained by us previously \cite{ks2}.

\subsection{$a,b,f$ = half-integers:}

We shall now discuss the case 
when three out of the four parameters (say $a,b,f$)
are half-integers while $g$ is any number (not a half-integer), 
and obtain the energies of the QES mid-band 
states. Our argument is again 
based on the assertion that 
relations (\ref{2.17}) and (\ref{2.18}) are simultaneously 
valid.
In particular, we assert that for
$a=k+\frac{1}{2},b=l+\frac{1}{2},f=n+\frac{1}{2}$ 
the three potentials
\bea\label{3.6}
&&[a=k+\frac{1}{2},b=l+\frac{1}{2},f=n+\frac{1}{2},g] \nonumber \\ 
&&\equiv \bigg [\frac{2(k+l+n+g)+5}{4},\frac{2(k+l-n-g)-1}{4},
\frac{2(k+n-l-g)-1}{4},\frac{2(l+n-k-g)-1}{4}\bigg] \nonumber \\ 
&& \equiv \bigg[\frac{2(k+l+n-g)+3}{4},\frac{2(k+l+g-n)+1}{4},
\frac{2(k+n+g-l)+1}{4}, \frac{2(l+n+g-k)+1}{4}\bigg]
\eea
have identical $k+l+n+3$ QES energy values.
What is being asserted here is that the $k+l+n+3$ 
QES mid-band energies of the potential with 
half-integer values of $a,b,f$ and arbitrary $g$ are the 
same as the band edge energy 
eigenvalues of the two other potentials in (\ref{3.6}).
This happens because for arbitrary values of
$g$, for the two potentials
\be\label{3.7}
\bigg[\frac{2(k+l+n+g)+5}{4},\frac{2(k+l-n-g)-1}{4},
\frac{2(k+n-l-g)-1}{4},\frac{2(l+n-k-g)-1}{4}\bigg]~, 
\ee
\be\label{3.8}
\bigg[\frac{2(k+l+n-g)+3}{4},\frac{2(k+l+g-n)+1}{4},
\frac{2(k+n+g-l)+1}{4}, \frac{2(l+n+g-k)+1}{4}\bigg]~,
\ee
we can always obtain
$k+l+n+3=a+b+f+3/2$ QES band edges, since for both
potentials the sum of the four numbers characterizing the potentials is
either $2k$ or $2l$ or $2n$. This is possible because the GAL
potential $[a,b,f,g]$ remains unchanged when any one (or more) 
of the four parameters $a,b,f,g$ change to
$-a-1,-b-1,-f-1,-g-1$ respectively. Thus we conjecture that the 
potential $[k+1/2,l+1/2,n+1/2,g]$, for arbitrary $g$ has $k+l+n+3$
QES mid-band states.
We now make a number of predictions for the energy
eigenvalues of the mid-band states for GAL potentials
of the form  $[k+1/2,l+1/2,n+1/2,g]$. 

\begin{enumerate}

\item
We predict that the potential $[1/2,1/2,1/2,g]$ has 
three QES mid-band states 
with energy values
\be\label{3.9}
E_1 = (1+m)~,~E_2=1+(g+1/2)^2 m~,~E_3=(g+1/2)^2+m~.
\ee

\item The potential $[3/2,1/2,1/2,g]$ 
has four QES mid-band states with energy values
\be\label{3.10}
E_1 = (4+m)~,~E_2=(g+1/2)^2+4m~,
~E_{3,4}= 2(1+m) \pm \sqrt{4(1-m)^2+(2g+1)^2 m}~.
\ee

\item The potential $[3/2,3/2,1/2,g]$ 
has five QES mid-band states with energy values
\bea\label{3.11}
&&E_1=(g+1/2)^2+4m~,~
E_{2,3}= 5+2m \pm \sqrt{4(2-m)^2+(2g+1)^2 m-4m}~, \nonumber \\
&&E_{4,5}= 5+[(g+1/2)^2 +1]m \pm \sqrt{[4-(2g+1)m]^2-(2g-3)^2 m+4m}~.
\eea

\item The potential $[3/2,3/2,3/2,g]$ 
has six
QES mid-band states with energy values
\bea\label{3.12}
&&E_{1,2}= (5+m) \pm \sqrt{16(1-m)^2+(2g+1)^2 m}~, \nonumber \\
&&E_{3,4}= 5+[(g+1/2)^2 +1]m \pm \sqrt{16-(2g+1)^2 m(1-m)}~, \nonumber
\\
&&E_{5,6}= [(g+1/2)^2 +1]+5m \pm \sqrt{16m^2+(2g+1)^2 (1-m)}~.
\eea

\item One can
readily obtain some QES mid-band energy values when
(i) $k$ is an arbitrary integer while
$l$ and/or $n$ are either $0$ or $1$ ; (ii) $k,l$ are arbitrary integers
with $n=0$ or $1$ and $g$ being any arbitrary number.
It would be nice to prove (or disprove) these conjectures and
more importantly, try to obtain the corresponding 
energy eigenfunctions. We shall have something to say about this point
when we discuss the implications of these results in the context of
Heun's equation. 

\item Note that
out of the $k+l+n+3$ QES states, the energy
values for $k+l+2$ states can also be obtained by considering
the previous case [$a=k+1/2,b=l+1/2$,$f,g=$arbitrary] and putting 
$f=n+1/2$ at the end of the calculation. As an illustration, 
consider the case $[3/2,1/2,f,g]$ where $f,g$
are arbitrary numbers.  As shown in the last subsection, there are
three QES energy values and they are given by eqs. (\ref{4.3a}) 
and (\ref{4.3b}). 
On putting $f=1/2$ in these equations,
we find that three (out of four) QES eigenvalues of
$[3/2,1/2,1/2,g]$ as given by eq. (\ref{3.11}) are correctly obtained.

\end{enumerate}

\subsection{\bf $a,b,f,g$ = half-integers:}

Finally, let us consider the case when all four parameters are 
half-integers. The potential is of the form $[k+1/2,l+1/2,n+1/2,p+1/2]$,
where $k,l,n,p$ are integers and we take 
$k \ge l \ge n \ge p$. 
We shall discuss the two cases when the sum $k+l+n+p$ is an even or an
odd integer separately.

\subsubsection{$k+l+n+p$ = even integer:}

As shown in I, for the GAL potential (\ref{1}), QES
energies are obtained when the sum of the four parameters is an
even integer including zero. 
In this case, as already 
shown by Takemura \cite{tak1}, both the
relations (\ref{2.17}) and (\ref{2.18}) are simultaneously valid.
We would like to assert here that in this case, in general there 
should be $k+n+l+p+4$ QES states, since 
the sum of the four numbers characterizing the potentials is
either $2k$ or $2l$ or $2n$ or $2p$. This is possible because the GAL
potential $[a,b,f,g]$ remains unchanged when any one (or more) 
of the four parameters $a,b,f,g$ change to
$-a-1,-b-1,-f-1,-g-1$ respectively. Unfortunately, in the several
specific cases that we have examined, we find that some of the
eigenvalues simply
get repeated. Thus the true number of
QES states may be much less than $k+l+n+p+4$. For example, 
consider the case
of $[5/2,1,2,1/2,1/2]$. While naively we expect six QES states, we only
find three QES levels at $E=(1+m),(1+9m),(9+m)$. Similarly, for the
potential $[9/2,1/2,1/2,1/2]$ while naively we expect 8 QES states, we
only find four QES levels at $E=(1+m),~(1+25m),~(25+m),~9(1+m)$.

\subsubsection{$k+l+n+p$ = odd integer:} 

Let us now discuss perhaps the most intriguing case when all four 
parameters are half-integers and their sum is an odd integer.
While it is clear from I that no QES band edges can be obtained when
$k+l+n+p$ is an odd integer, it is not obvious whether mid-band states
can be obtained in this case. 
In fact we shall now obtain energy values for several QES mid-band
states for these potentials. As argued in the introduction, while
Takemura has shown that relations (\ref{2.17}) or (\ref{2.18}) are
valid when $a+b+f+g$ is an even or an odd integer respectively (and
all are integers), we conjecture that irrespective of whether $a+b+f+g$
is an odd or an even integer, both the relations are always valid.
Using eqs. (\ref{2.17}) and (\ref{2.18}), it then follows that 
irrespective of whether $a+b+f+g$ is an even or an odd integer, they 
always have a partner potential of the form $[k+1/2,l+1/2,n+1/2,p+1/2]$
where $k,l,n,p$ are all integers and their sum is an odd integer.
We thus conjecture that when $a+b+f+g$ is an
even integer, then the QES mid-band energy values of the potential
$[(a+b+f+g+1)/2,(a+b-f-g-1)/2,(a+f-b-g-1)/2,(a+g-b-f-1)/2]$
are the same as the band edge eigenvalues of the potential $[a,b,f,g]$.
Similarly, when $a+b+f+g$ is an
odd integer, then we conjecture that 
the QES mid-band energy values of the potential
$[(a+b+f-g)/2,(a+b+g-f)/2,(a+f+g-b)/2,(b+f+g-a)/2]$
are the same as the band edge eigenvalues of the potential $[a,b,f,g]$.
This is rather remarkable.  
Since the band edges of the GAL potentials $[a,b,f,g]$ with integer
$a,b,f,g$ are all known (at least in principle), hence 
one has a prediction for $k+l+n+p+4$ QES mid-band states of the GAL potentials
of the type $[k+1/2,l+1/2,n+1/2,p+1/2]$ when $k+l+n+p$ is an 
odd integer. 
To be precise, we predict that the potential $[k+1/2,l+1/2,n+1/2,p+1/2]$
has $k+l+n+p+4$ QES mid-band states in case $k+l+n+p$ is an odd integer
and these eigenvalues are identical to the band edges of the potential
$[(k+l+n+p+3)/2,(k+l-n-p-1)/2,(k+n-l-p-1)/2, (k+p-l-n-1)/2]$. 
As an illustration, knowing the band edges of the Lam\'e
potential $2m\sn^2(x)$ we predict that the three QES mid-band
state energy values of the potential
$[1/2,1/2,1/2,-1/2]$ must be at $E=m,1,1+m$. Similarly, we predict that
the five mid-band QES energies of the potential
$[3/2,1/2,1/2,1/2]$ are at
\be\label{4.9}
E_2=1+m~,~E_3=(1+4m)~,~E_4 =(4+m)~,
~E_{1,5}=2(1+m) \pm 2\sqrt{1-m+m^2}~,
\ee

The validity of our conjecture and in a way the consistency of our whole
approach can be checked by extrapolating the results obtained in the
previous subsection. In particular, in the last subsection we
have discussed the case when the potential is of the form
$[k+1/2,l+1/2,n+1/2,g]$ where $g$ is any arbitrary number 
and have seen
that in that case one obtains energies for 
$k+l+n+3$ QES mid-band states. On choosing $g=p+1/2$ and choosing 
$k,l,n,p$ such that their sum is an odd integer, we have verified in many
cases the validity of our conjectures.

We thus predict that the potentials 
$[3/2,1/2,1/2,1/2]$, $[5/2,1/2,1/2,-1/2]$ and $[3/2,3/2,1/2,-1/2]$ 
have exactly the same (five) 
QES mid-band energies as the band edge energy eigenvalues of the 
potentials $[2,0,0,0],~[2,1,1,0],~[2,1,0,0]$ respectively. 
It is worth observing that
there are exactly three ways of obtaining $k+l+n+p=1$ given that 
$k \ge l \ge n \ge p$ and that while $k,l,n$ are non-negative
integers, $p$ is $\ge -1$. It may be noted here that in case $n$ is
also $-1$ then the corresponding partner potentials as given by eqs.
(\ref{2.17}) and (\ref{2.18}) are potentials of the type
$[a,a,b.b]$ and similarly if both $n=l=-1$ then the partner potentials
are of type $[a,a,a,a]$ which as explained in Sec. 2,
 are not included in our analysis. Thus, put another way, the problem
of finding the number of
independent potentials with say $a$ band gaps reduces to finding 
four integers $k,l,n,p$ with 
$k \ge l \ge n \ge p$ (with $n \ge 0,p \ge -1$)   
such that their sum equals $2a-3$. 

\section{\bf GAL Potentials [$a,b,f,g$ = arbitrary numbers].}

So far, we have discussed the 
cases when the four parameters $a,b,f,g$
take integer values or at least one of them is half-integer.
Now we want to extend this
discussion to the general case when $a,b,f,g$ 
take arbitrary values.

As seen in Sec. 4, when either one, two or 
three of the parameters are
half-integer while the remaining parameters are arbitrary, 
then clearly the corresponding partners indeed correspond to the case
where $a,b,f,g$ are arbitrary numbers. We therefore conjecture that
 the relations (\ref{2.17})
and (\ref{2.18}) are valid even when the four parameters $a,b,f,g$ take any
arbitrary values, the only restriction being that 
either $a+b+f+g$ or any other combination obtained by replacing one
or more of these parameters by $-a-1,-b-1,-f-1,-g-1$ respectively 
is a non-negative even integer.
Further,
 in that case the two partner potentials have the same band-edge eigenvalues
as the QES mid-band energies of the 
potentials where either one, two or
three of the parameters are half-integers.

A few illustrative examples are in order here. 
Consider the potential $[4/5,2/5,2/5,2/5]$. In this case
while $a+b+f+g=2$, no other combination characterizing 
relations (\ref{22.2}) to (\ref{22.8a}) gives an even 
integer. Using eq.
(\ref{2.17}) we find 
that this potential has a GAL partner $[3/5,3/5,3/5,1/5]$. Using Table 4 
of I it is easily shown
 that both these potentials 
 have two (identical) QES band edge energy 
eigenvalues 
\be\label{6.1}
E= \frac{26}{25}(1+m)\pm \frac{2}{5}\sqrt{1-m+m^2}~.
\ee

If instead we consider the potential $[17/5,8/5,7/5,6/5]$, then only
$a+f-b-g$ is an integer and the corresponding GAL partner
potential is $[13/5,12/5,11/5,2/5]$ and both have one (identical) QES
energy. Of course it can happen that $a,b,f,g$ are
such that more than one of the relations (\ref{22.1}) to (\ref{22.8a}) 
are satisfied. In that case one has more QES
band edge eigenvalues. For example, if the potential parameters are such that 
$a+b-g-f-1,a+f-b-g-1$ as well as $a+g-f-b-1$ are nonnegative 
integers, then using eqs. (\ref{22.2})
to (\ref{22.4}) it is easily seen that the number of QES energy
eigenvalues is equal to $(3a-b-f-g)/2$. 
One illustration of this is the potential
$[11/5,1/5,1/5,1/5]$ which has three QES energies 
$(1+m),144/25+m,1+(144/25)m$. We might add here that the discussion
above is valid even if the numbers $a,b,f,g$ are irrational numbers 
but such that $a+b+f+g$ or any other combination obtained by replacing
one or more of $a,b,f,g$ to $-a-1,-b-1,-f-1,-g-1$ respectively is an
even integer (including zero).

We would like to restate here that when all four parameters
$a,b,f,g$ are integers, one has a finite number of band gaps. In all
other cases one expects to have an infinite number of bands and
band gaps out of which only a few are QES states.

\section{\bf Implications for Heun's Equation.}

Heun's equation, a second order linear differential
equation with four regular singular points has been extensively
discussed in the mathematics literature \cite{ron,mai,erd}. 
The intimate connection between Heun's equation and GAL potentials 
is well known \cite{spa}. 
In recent years, this 
equation has also proved very useful in the context of a number of
physical problems, like quasi-exactly solvable systems \cite{jnk},
sphaleron stability \cite{bre}, Calogero-Sutherland models 
\cite{tak2},
higher dimensional correlated systems \cite{bklms}, Kerr-de Sitter black
holes \cite{stu}, and finite lattice Bethe ansatz systems \cite{dst}.

The canonical form of Heun's equation is given by \cite{ron}
\be\label{5.1}
\bigg [\frac{d^2}{dx^2}+\big (\frac{\gamma}{x}+\frac{\delta}{x-1}
+\frac{\epsilon}{x-c} \big )\frac{d}{dx}+\frac{\alpha \beta
x-q}{x(x-1)(x-c)} \bigg ]G(x) =0~,
\ee
where $\alpha,\beta,\gamma,\delta,\epsilon,q,c$ are parameters, except
that $c \ne 0,1$ and the first five parameters are related by
\be\label{5.1a}
\gamma+\delta+\epsilon=\alpha+\beta+1~.
\ee
The four regular singular points of eq. (\ref{5.1}) 
are located at $x=0,1,c$ and the point at infinity.

If we make the transformation $x=\sn^2(y,m)$, then Heun's equation takes
the form  \cite{ron} 
\bea\label{5.2}
&&F''(y)+[(1-2\epsilon) m \frac{\sn(y,m) \cn(y,m)}{\dn(y,m)}
+(1-2\delta)
\frac{\sn(y,m)\dn(y,m)}{\cn(y,m)} +(2\gamma-1) \frac{\cn(y,m)
\dn(y,m)}{\sn(y,m)}]F'(y) \nonumber \\
&&-[4mq
-4\alpha \beta m \sn^2 (y,m)]F(y) =0~,
\eea
where $c=1/m$, $G(x) \equiv F(y)$.
The periodic solutions of eq. (\ref{5.2}) correspond to the 
polynomial solutions of eq. (\ref{5.1}) while the quasi-periodic solutions 
correspond to non-polynomial solutions of (\ref{5.1}).

The interesting point is that after a transformation, the
Schr\"odinger equation (\ref{2.1}) for the GAL potential
(\ref{1}) is in fact Heun's eq. (\ref{5.2}). 
In particular, let us start from the Schr\"odinger equation (\ref{2.1})
for the GAL potential (\ref{1}). On substituting
\be\label{5.6}
\psi(y)=\dn^{-b}(y)\cn^{-f}(y)\sn^{-g}(y)\phi(y)~,
\ee
one can show that $\phi(y)$ satisfies the differential
equation
\bea\label{5.7}
&&\phi''(y)+2[mb \frac{\sn(y,m) \cn(y,m)}{\dn(y,m)}
+f\frac{\sn(y,m)\dn(y,m)}{\cn(y,m)} -g \frac{\cn(y,m)
\dn(y,m)}{\sn(y,m)}]\phi'(y) \nonumber \\
&&-[R
-Q m \sn^2 (y,m)]\phi(y) =0~,
\eea
where 
\be\label{5.8}
R=-E+m(g+b)^2 +(f+g)^2~,~~Q=(b+f+g)(b+f+g-1)-a(a+1)~.
\ee
Thus once we obtain solutions of the Schr\"odinger equation
for the GAL potential (\ref{1}), then we can immediately write the
solutions for the periodic form of Heun's eq. (\ref{5.2}) and the 
solutions of the original Heun's eq. (\ref{5.1}) with the identification
\bea\label{5.9}
&&\gamma = \frac{1}{2} -g~,~\delta = \frac{1}{2} -f~,~\epsilon \frac{1}{2}-b~, \nonumber \\
&&\alpha+\beta =\frac{1}{2}-(b+f+g)~,~4\alpha \beta =Q~,~4mq=R~,
F(y) \equiv \phi(y)~.
\eea

We now make a crucial observation. From eq. (\ref{5.8}), it follows 
that if under any transformation, the parameters $b_1,f_1,g_1$ 
change to $b_2,f_2,g_2$ and the energy $E$ remains invariant, then  
the corresponding  values of $R$ are related by
\be\label{5.10}
R_1 -m(b_1+g_1)^2-(f_1+g_1)^2=R_2 -m(b_2+g_2)^2-(f_2+g_2)^2~.  
\ee
Making use of eq. (\ref{5.10}) and the connection between GAL
potentials as given by eqs. (\ref{2.17}) and (\ref{2.18}), we can 
obtain interesting relations for Heun's equation.
Using the fact that the two GAL potentials given by eq.
(\ref{2.17}) have the same band edge energy eigenvalues and the 
eigenfunctions for both the partners as are given by 
eqs. (\ref{22.1}) to (\ref{22.4a}), 
one can obtain the connection between the 
two corresponding solutions of Heun's equation. 
For example, consider the solution (\ref{22.1}) and the corresponding
solution of the GAL partner obtained by the above substitution.
Using eq. (\ref{5.10}) it then follows that corresponding to a 
given periodic (i.e. polynomial) solution of Heun's equation with
parameter set ($\alpha, \beta, \delta, \epsilon,\gamma,q$) there
always exists another periodic solution 
with the {\it same} $q$ provided the
other parameters change as follows:
\be\label{5.11}
\gamma \rightarrow \alpha~,~\alpha \rightarrow \gamma~,~\beta 
\rightarrow \beta~,~\epsilon \rightarrow 1+\beta-\delta~, 
\delta \rightarrow 1+\beta-\epsilon~.
\ee
Similarly, on considering the other three periodic solutions as given by eqs.
(\ref{22.2}) to (\ref{22.4}) and the corresponding solutions 
of the GAL partner potential with the same energy, we find that corresponding
to a given periodic solution of Heun's equation,
there exist the following three (periodic) solutions with the 
change of parameters
given by (note that $R=4mq$ and $c=1/m$)
\be\label{5.12}
\gamma \rightarrow 1+\beta -\epsilon~,~\alpha \rightarrow \delta~,~\beta 
\rightarrow \beta~,~\epsilon \rightarrow 1+\beta-\gamma~, 
\delta \rightarrow \alpha~,~q \rightarrow q- \beta (\delta-\alpha)
\ee
\be\label{5.13}
\epsilon \rightarrow \alpha~,~\alpha \rightarrow \epsilon~,~\beta 
\rightarrow \beta~,~\gamma \rightarrow 1+\beta-\delta~, 
\delta \rightarrow 1+\beta-\gamma~,~q \rightarrow
q-\beta (\epsilon-\alpha)c~,
\ee
\be\label{5.14}
\gamma \rightarrow 1+\beta-\gamma~,~\alpha \rightarrow 1+\beta -\alpha~,
~\beta 
\rightarrow \beta~,~\epsilon \rightarrow 1+\beta-\epsilon~, 
\delta \rightarrow 1+\beta-\delta~,q \rightarrow q+\beta 
[(\alpha-\delta) +(\alpha-\epsilon)c]~.
\ee

Thus given a periodic solution of Heun's equation, one immediately 
has 4 other periodic solutions as given by eqs. (\ref{5.11}) to (\ref{5.14}). 
We have checked that if instead we consider the two partner GAL
potentials given by eq. (\ref{2.18}) and consider the corresponding
eigenfunctions given by eqs. (\ref{22.5}) to (\ref{22.8a}) 
(and those 
of the corresponding GAL partner potentials with the same energy), then we
again obtain the {\it same} relations [eqs. (\ref{5.11}) to
(\ref{5.14})]. As an additional check,
we have looked at the partner GAL potentials as given by eqs.
(\ref{3.1}), (\ref{3.7}), (\ref{3.8}), (\ref{4.1}) and
(\ref{4.6}) and in all these
cases we get back the relations (\ref{5.11}) to (\ref{5.14}), which 
to the best of our knowledge, are new results.
Several comments are in order:

\begin{enumerate}

\item In Sec. 4 we have shown that the QES mid-band energy values
of the GAL
potential $[a=k+1/2,b,f,g]$ are the same as the QES 
energies of the two 
GAL potentials given in (\ref{3.1}). 
What does this imply in
the context of Heun's equation? It is easily shown that as a consequence
of the discussion in Sec. 4,
given a periodic solution of Heun's equation with the set of parameters
$\alpha,\beta,\gamma,\delta,\epsilon,q$, one has a 
corresponding quasi-periodic solution with changed parameters:
\bea\label{5.15}
&&\gamma \rightarrow 2-\alpha~,~\alpha \rightarrow 1+\gamma -\alpha~,
~\beta 
\rightarrow 1+\beta-\alpha~,~\epsilon \rightarrow 1+\beta-\delta~, 
\delta \rightarrow 1+\beta-\epsilon~, \nonumber \\
&&q \rightarrow q+(\alpha -1)
[(1+\beta-\delta) +(1+\beta-\epsilon)c]~.
\eea
An additional check on this relation is obtained  by using the connection 
between the mid-band states of the GAL
potentials $[a=k+1/2,b=l+1/2,f,g]$ and 
$[a=k+1/2,b=l+1/2,f=n+1/2,g]$ and the QES energies of the
potentials (\ref{4.1}), (\ref{3.7}) and (\ref{3.8}) 
and we again obtain the {\it same}
connection between the periodic and quasi-periodic solutions of 
Heun's equation.

\item Using the results
in Sec. 4 regarding the case when three of the four parameters
$a,b,f,g$ are half-integers, we also obtain two more relations 
connecting the periodic and quasi-periodic solutions of Heun's
equation. In particular, given a periodic solution of Heun's equation
with the set of parameters
$\alpha,\beta,\gamma,\delta,\epsilon,q$, it implies the following two
quasi-periodic solutions of Heun's equation:
\be\label{5.16}
\gamma \rightarrow \alpha~,~\alpha \rightarrow 1+\alpha -\epsilon~,
~\beta 
\rightarrow \delta~,~\epsilon \rightarrow 1+\delta-\beta~, 
\delta \rightarrow 1+\beta-\epsilon~,q \rightarrow q+ \alpha
(\beta-\delta)~,
\ee
\be\label{5.17}
\gamma \rightarrow \alpha~,~\alpha \rightarrow 1+\alpha -\delta~,
~\beta 
\rightarrow \epsilon~,~\epsilon \rightarrow 1+\beta-\delta~, 
\delta \rightarrow 1+\epsilon-\beta~,q \rightarrow q+4\alpha 
(\beta-\epsilon)c~.
\ee

\item Needless to say that if instead, a quasi-periodic solution of Heun's
equation is given, then by inverting eqs. (\ref{5.15}) to (\ref{5.17}),
we immediately obtain three periodic solutions of Heun's equation.  
\end{enumerate}

So far we have discussed how the connections between different GAL
potentials can help in finding new solutions of Heun's equation. It
may happen that in some cases it may be simpler to solve the 
algebraic Heun's equation (\ref{5.1}) rather than its periodic 
variant. We now show that this is indeed so in the case of several
quasi-periodic mid-band eigenfunctions. Consider for example the
GAL potentials when either $b$ or $f$ or $g$ is 1/2 while the other 
three parameters are arbitrary. Using arguments of
section IV, we can easily obtain the eigenvalues for mid-band 
states for these potentials.
We shall now show that using these
eigenvalues we can easily solve the algebraic form of 
Heun's eq. (\ref{5.1}) and hence
using the connection as explained above, obtain the 
eigenfunctions for the mid-band states of these GAL potentials.

Consider the 
GAL potential $[a,1/2,f,g]$ where $a,f,g$ are arbitrary numbers,
Note that using the relations (\ref{2.17}), (\ref{2.18}) and
Table 4 of I, it is easily shown that the QES 
mid-band eigenvalue 
of the GAL potential $[a,1/2,f,g]$ is at   
$E=(a+1/2)^2+m(g+1/2)^2$.
On using the connection formulas (\ref{5.8}) and (\ref{5.9}) 
it is easily shown that the corresponding parameters 
for Heun's eq. (\ref{5.1}) are
$$
\gamma=1/2 -g~,~\delta=1/2-f~,~\epsilon = 0~,~\alpha=(a-f-g+1/2)/2~,
$$
\be\label{5z}
~\beta=-(a+f+g+1/2)/2~,~q=(a+f+g+1/2)(f+g-a-1/2)(c/4)~.
\ee
Remarkably, for these parameters, it is straightforward to obtain the
solution of the algebraic Heun's eq. (\ref{5.1}) and show that
\be\label{5y}
G(x) = F[(a-f-g+1/2)/2,-(a+f+g+1/2)/2,1/2-g;x]~,
\ee
where $F(a,b,c;x)$ is the hypergeometric function.
The corresponding mid-band state eigenfunction for the GAL potential
 $[a,1/2,f,g]$ is then immediately written down. We have verified that
 this is indeed the correct eigenfunction in the following cases (i)
 $a=f=g=0$ (ii) $a$ integral, $f=g=0$ (iii) $a$ arbitrary 
 while $f,g$ are integral.

Similarly, for the GAL potential $[a,b,1/2,g]$, with arbitrary $a,b,g$,
the QES mid-band energy eigenvalue is $E=(g+1/2)^2 +(a+1/2)^2 m$ and
proceeding as above, it is easily shown that the solution of the 
algebraic Heun's eq. (\ref{5.1}) is given by
\be\label{5yy}
G(x) = F[(a-b-g+1/2)/2,-(a+b+g+1/2)/2,1/2-g;mx]~.
\ee

And finally, for the GAL potential $[a,b,f,1/2]$, with arbitrary $a,b,f$,
the QES energy is $E=(f+1/2)^2 +(b+1/2)^2 m$ and
proceeding as above, it is easily shown that the solution of the 
algebraic Heun's eq. (\ref{5.1}) is given by
\be\label{5yz}
G(x) = F[(a-b-f+1/2)/2,-(a+b+f+1/2)/2,1/2-b;(1-mx)/(1-m)]~.
\ee
On using the solutions (\ref{5y}) to (\ref{5yz}) it follows that for the
potential $[a,1/2,1/2,1/2]$ one knows three QES mid-band energy
eigenstates. In the special case when $a=2k+3/2$ these eigenstates are
the mid-band QES eigenstates for the potential $[2k+3/2,1/2,1/2,1/2]$
where the sum of the four parameters characterizing the potential is an
odd integer.

\section{\bf Summary and Open Questions.}

In this paper, we have addressed many issues regarding GAL
potentials with a number of choices for the parameters $a,b,f,g.$ The most
interesting case is when all the four parameters are integers. This is
a potential with a finite number of band gaps. We have been able to count the 
number of independent GAL potentials with a given number of band gaps and 
completely specify the nature of the band
edge eigenfunctions. We have introduced the new
concept of self-dual potentials which are not self-isospectral. We are
also able to specify how many of the independent potentials with a given
number of band gaps have supersymmetric partner potentials and how many
have non-supersymmetric partner potentials. Finally, using the results
for the GAL potentials, we have shown that given any one periodic solution of
Heun's equation, one can obtain four more periodic solutions. 

We have also discussed several issues related with GAL potentials when 
one or more of the parameters take half-integer values. In
particular, while nothing is known so far
about GAL potentials when three of
the parameters take half-integer values, we have been able to obtain 
the QES energy values for several of these potentials.
Further, using these eigenvalues and the algebraic form of Heun's
equation, we have also been able to obtain the corresponding
eigenfunctions for potentials of the form $[a,1/2,f,g],[a,b,1/2,g],
[a,b,f,1/2]$ where $a,b,f,g$ are arbitrary numbers. 
The key point to make while addressing these questions is that
the relations (\ref{2.17}) and (\ref{2.18}) are not only valid when the
four parameters $a,b,f,g$ are integers but also when one or more of these
parameters take half-integer values.
This in turn immediately implies that these relations are also valid when
the four parameters $a,b,f,g$ take arbitrary values so long as either 
their sum $a+b+f+g$ [or one or more of the combinations obtained by changing
one or more of the parameters to $-a-1,-b-1,-f-1,-g-1$ respectively] is a
nonnegative even integer. We have also conjectured that both
relations (\ref{2.17}) and (\ref{2.18}) are simultaneously valid 
when $a,b,f,g$ are integers and that the energy eigenvalues for the 
band edges of these potentials are the same as mid-band QES energy values
of GAL potentials in which all 
four parameters are half-integers and their sum is an odd integer. 
Finally, using these results we have also shown that given a 
periodic solution of Heun's equation, one can immediately obtain
three quasi-periodic solutions of the same equation. 

This work raises several issues which we have not been able to
address satisfactorily:

\begin{enumerate}

\item Can one explicitly write down all seven KdV equations of seventh
order?

\item What are the QES 
eigenfunctions for GAL potentials when one or more of the parameters is
half-integral ($\ge 3/2$) while the remaining parameters are arbitrary?

\item The problem when two of the four parameters are half-integers needs 
further study. In particular, it is still not clear how many QES energy
eigenvalues can be obtained, in general, 
in that case. 

\item When the
sum of all the four parameters is an even integer, it is clear that the QES
states correspond to band edges. However, a complete understanding is still
lacking regarding the number of QES states for various values of $a,b,f,g$. 
Further, when the sum of the four parameters is an odd
integer, the form of the QES eigenfunctions is not clear when the
half-integer parameters are $>1/2$.

\end{enumerate}

\end{document}